\documentclass[aps,
twocolumn,nofootinbib,nobibnotes,preprintnumbers]{revtex4-1}

\usepackage{url,comment}
\usepackage{times}
\usepackage{latexsym}
\usepackage{graphicx, graphics, hyperref, amsmath, amssymb, slashed, xcolor, bbm, amsthm, array, multirow, tabularx, tabulary,  makecell}
\usepackage{subfigure}
\usepackage{rotating}
\usepackage{afterpage}
\newcommand{\nc}{\newcommand}
\nc{\beq}{\begin{equation}}
\nc{\eeq}{\end{equation}}
\nc{\barray}{\begin{eqnarray}}
\nc{\earray}{\end{eqnarray}}
\nc{\barrayn}{\begin{eqnarray*}}
\nc{\earrayn}{\end{eqnarray*}}
\nc{\bcenter}{\begin{center}}
\nc{\ecenter}{\end{center}}
\nc{\mc}{\mathcal}
\nc{\er}[1]{(\ref{eq:generator_level_cuts_on_jets}\ref{eq:#1})}
\nc{\onehalf}{\frac{1}{2}} 
\nc{\partialbar}{\bar{\partial}}
\nc{\psit}{\widetilde{\psi}}
\nc{\Tr}{\mbox{Tr}}
\nc{\hc}{\mbox{H.c.}}
\nc{\ev}{\;\mathrm{eV}}
\nc{\mev}{\;\mathrm{MeV}}
\nc{\gev}{\;\mathrm{GeV}}
\nc{\tev}{\;\mathrm{TeV}}

\def\chii0{\chi_i^0}
\def\chij0{\chi_j^0}

\newcommand{\gsim}{\lower.7ex\hbox{$\;\stackrel{\textstyle>}{\sim}\;$}}
\newcommand{\lsim}{\lower.7ex\hbox{$\;\stackrel{\textstyle<}{\sim}\;$}}
\nc{\ttbar}{t\bar t}

\newcommand{\cref}[1]{Chapter~\ref{c.#1}}

\graphicspath{{figs/}}


\begin{document}
\preprint{TTP18-043}
\preprint{PP-018-10} 
\title{
Closing the light gluino gap with electron-proton colliders
}

\author{David Curtin}
\email{dcurtin@physics.utoronto.ca}
\affiliation{Department of Physics,\\ University of Toronto, Canada}

\author{Kaustubh Deshpande}
\email{ksd@umd.edu}
\affiliation{Maryland Center for Fundamental Physics, Department of Physics,\\ University of Maryland, College Park, MD 20742-4111 USA}

\author{Oliver Fischer}
\email{oliver.fischer@kit.edu}
\affiliation{Institute for Nuclear Physics (IKP), Karlsruhe Institute of Technology, 
Hermann-von-Helmholtz-Platz 1, D-76344 Eggenstein-Leopoldshafen, Germany}

\author{Jos\'e Zurita }
\email{jose.zurita@kit.edu}
\affiliation{
\mbox{Institute for Nuclear Physics (IKP), Karlsruhe Institute of Technology,}\\
\mbox{Hermann-von-Helmholtz-Platz 1, D-76344 Eggenstein-Leopoldshafen, Germany}\\ \phantom{a}\vspace*{-2mm} \\
\mbox{Institute for Theoretical Particle Physics (TTP), Karlsruhe Institute of Technology,}\\ 
\mbox{Engesserstra{\ss}e 7, D-76128 Karlsruhe, Germany} \vspace*{2mm}
}

\date{\today}
\begin{abstract}

The future electron-proton collider proposals, LHeC and FCC-he, can deliver $\mathcal{O}$(TeV) center-of-mass energy collisions, higher than most of the proposed lepton accelerators, with $\mathcal{O}$(ab$^{-1}$) luminosity, while maintaining a much cleaner experimental environment as compared to the hadron machines. This unique capability of $e^- p$ colliders can be harnessed in probing BSM scenarios giving final states that look like hadronic noise at $pp$ machines. In the present study, we explore the prospects of detecting such a prompt signal having multiple soft jets at the LHeC. Such a signal can come from the decay of gluino in RPV or Stealth SUSY, where there exists a gap in the current experimental search with $m_{\tilde{g}} \approx 50 - 70$ GeV. We perform a simple analysis to demonstrate that, with simple signal selection cuts, we can close this gap at the LHeC at 95\% confidence level, even in the presence of a reasonable systematic error. More sophisticated signal selection strategies and detailed knowledge of the detector can be used to improve the prospects of signal detection.

\end{abstract}

\maketitle

\section{Introduction}
\label{s.intro}

The search for new physics which can address fundamental problems in the Standard Model (SM) like the hierarchy problem, dark matter, baryon asymmetry etc., depends crucially on the development of experiments capable of probing new regimes. Colliders searching for physics Beyond the Standard Model (BSM) form a major and important part of such experiments. Hadron or $pp$ colliders, the Large Hadron Collider (LHC) or the future colliders  FCC-hh \cite{Golling:2016gvc, Mangano:2016jyj, Contino:2016spe} and SppC \cite{Tang:2015qga}, with their enormous center-of-mass energy and luminosity, provide increasing capability to probe physics at higher mass scales, or very rare processes. The future proposed lepton or $e^+ e^-$ colliders like ILC \cite{Baer:2013cma, Brau:2015ppa}, FCC-ee \cite{Gomez-Ceballos:2013zzn}, CEPC \cite{CEPCStudyGroup:2018rmc, CEPCStudyGroup:2018ghi}, and CLIC \cite{Aicheler:2012bya}, although with a much lower center-of-mass energy than the $pp$ colliders, due to very low backgrounds, are ideal for precision measurements, now particularly in the case of Higgs physics.

The electron-proton ($e^- p$) colliders are a hybrid of $e^+ e^-$ and $pp$ machines.  HERA, the only $e^- p$ collider built so far, via deep inelastic scattering (DIS) measurements, provided information about the Parton Distribution Functions (PDFs). However, due to its lower center-of-mass energy (320 GeV) and luminosity ($\sim$500 pb$^{-1}$), it was outclassed in almost all BSM searches by Tevatron (1.96 TeV and 10 fb$^{-1}$), except in the case of specific leptoquark scenarios \cite{South:2013fta, Abazov:2011qj}. The future proposals for $e^- p$ colliders, LHeC \cite{Klein:2009qt, AbelleiraFernandez:2012cc, Bruening:2013bga} and FCC-he \cite{Zimmermann:2014qxa}, are thought of as an add-on or upgrade to the corresponding $pp$ counterpart, which hence can be installed at a much lower cost (almost an order of magnitude less) than that of the $pp$ collider, while providing invaluable information on the structure of the proton and the Higgs and top sectors.

The unique feature of these future $e^- p$ colliders is that their proposed center-of-mass energy of 1.3 (3.5) TeV at the LHeC (FCC-he) is much higher than that of most of the proposed lepton colliders, along with $\sim$1 ab$^{-1}$ total luminosity, while maintaining a clean experimental environmental with much less QCD backgrounds and pile-up as compared to the hadron colliders. Thus, apart from doing more precise measurements of PDFs, this capability of $e^- p$ colliders can also be harnessed in searches for new physics. This can be particularly useful in probing BSM scenarios which provide signal final states that look like hadronic noise at the $pp$ colliders, thus making them very hard to probe. Ref.~\cite{Curtin:2017bxr} studied such ``stealth'' signals that arise in the form of long-lived particles, which can be an essential component of various BSM scenarios and hence theoretically very well motivated. (See \cite{Curtin:2018mvb} for a review.) However, these ``stealth'' signals can also be prompt. Soft multi-jet final states with no appreciable missing transverse energy (MET) at $pp$ colliders, which we will focus on in the current study, is a prime example of this.

In Section~\ref{s.lightgluinos}, we briefly review how such a soft multi-jet signal can arise from the gluino decay in R-parity violating (RPV) or Stealth supersymmetry (SUSY). Here we also describe the existing gap in the experimental search for such light gluinos. In Section~\ref{s.epbasics}, we review the salient features of $e^- p$ colliders, especially the LHeC. We study light RPV gluinos at the LHeC in Section~\ref{s.gluinosatep}. We find that in the clean environment of LHeC, the soft multi-jet signal from these light gluinos can be observed at 95\% confidence level, even in the presence of a reasonable systematic error, by making use of simple signal selection cuts. Thus, the existing gap in the search of these light gluinos can be closed with LHeC. We conclude in Section~\ref{s.conclusions}.

\section{Light Gluinos}
\label{s.lightgluinos}

Gluinos are a prediction of supersymmetric models~\cite{Martin:1997ns} and are expected to have a mass below a few TeV if the little hierarchy problem is addressed by weak- or TeV-scale stops in the spectrum. The LHC results exclude the possibility of gluino with mass $\lesssim$ 2 TeV, for particular decay paths for gluino decaying to multiple jets with missing energy signal \cite{Aaboud:2017vwy, Aaboud:2018mna}. More general analysis of the LHC data allowing for almost any decay path puts this lower limit at $\sim$1 TeV \cite{Evans:2013jna}. The MSSM gluino is bound to decay into SM particles and missing energy due to the R-parity.

However, if the gluino has all-hadronic multi-jet decay with no appreciable missing transverse energy, then the above bounds do not apply. Such a decay, e.g. $\tilde{g} \rightarrow jjj$, can come from R-parity violating (RPV) \cite{Barbier:2004ez} or Stealth SUSY \cite{Fan:2011yu} models. In the RPV SUSY models, this decay can happen via intermediate heavy squark decaying to two quarks (via the RPV coupling) with no invisible particle in the final state: $\tilde{g} \rightarrow q \tilde{q} \rightarrow q q q$. In the case of Stealth SUSY, the lightest supersymmetric particle (gravitino or axino) can carry away very small invisible energy, possibly below the detector resolution. This is due to the near-degenerate mass spectrum of the new weak-scale colored particles $(S, \tilde{S})$. Thus the gluino decay here, $\tilde{g} \rightarrow g \tilde{S} \rightarrow g S \tilde{G} \rightarrow g j j \tilde{G}$, can mimic a prompt $\tilde{g} \rightarrow jjj$ decay.

If the gluino is very light, along with having a prompt decay to multiple jets without any MET, this soft multi-jet signal suffers from very large QCD backgrounds and is extremely challenging to trigger on at $pp$ colliders. Indeed, the bounds provided by LHC for such gluinos apply only for $m_{\tilde{g}} \geq 100$ GeV \cite{ATLAS:2012dp, Chatrchyan:2011cj, Chatrchyan:2012uxa} while the lowest mass reach of $m_{\tilde{g}} = 77$ GeV is provided by CDF detector at Tevatron \cite{Aaltonen:2011sg}. Furthermore, the analysis of LEP event shape data excludes $m_{\tilde{g}} \leq 51$ GeV, independent of the decays of the gluino \cite{Kaplan:2008pt}. This leaves a gap in the search for these light RPV gluinos for $m_{\tilde{g}}$ between 51 and 76 GeV. As shown in \cite{Evans:2018scg}, even the data from UA2 detector at the SPS collider is not able to close this gap. 

\section{Electron-proton collider basics}
\label{s.epbasics}

In this section, we briefly review the important features of the proposed electron-proton colliders, which are relevant for the current study. Electron-proton colliders are hybrids of $e^+ e^-$ and $pp$ colliders. Although they are usually considered only in the context of deep inelastic scattering measurements, they can also provide a unique advantage over $e^+ e^-$ and $pp$ colliders in many searches of beyond the standard model (BSM) physics, see for instance \cite{Azuelos:2018syu}. 
The Large Hadron electron Collider (LHeC) \cite{Klein:2009qt, AbelleiraFernandez:2012cc, Bruening:2013bga} is one of the future proposed $e^- p$ colliders at CERN. It will have a 7 TeV proton beam of the high-luminosity LHC colliding with a 60 GeV electron beam generated from a separate electron accelerator (linac). Running concurrently with the HL-LHC, LHeC will have a center-of-mass energy of $\sim$1.3 TeV with the total luminosity of $\sim$1  $ab^{-1}$ over it's entire lifetime of $\sim$10 years. At the collision point in the HL-LHC tunnel, the LHeC detector will be installed. It will be a general-purpose detector accommodating for the asymmetric nature of the $e^- p$ collision. The $p_T$ threshold to reconstruct jets at LHeC is as small as 10 GeV, provided that the total energy is above 20 GeV~\cite{AbelleiraFernandez:2012cc}.

The proposal for the Future Circular Collider (FCC) at CERN also includes the Future Circular hadron-electron Collider (FCC-he) \cite{Zimmermann:2014qxa} which will be an even more powerful $e^- p$ collider. Here, the LHeC electron beam will collide with the 50 TeV FCC proton beam forming a collision with center-of-mass energy of about 3.5 TeV and total luminosity of about 1 ab$^{-1}$. In this work, we will focus on the analysis at the LHeC since it can be installed as early as 2023, while the FCC may operate in a more distant future.

The LHeC will have very low pile-up and QCD background as compared to the LHC. Thus it can potentially provide a unique advantage in the searches of BSM scenarios which give rise to final states with multiple soft jets that have no distinctive feature (e.g: no leptons, little missing energy, no activity in the muon system) from pure QCD processes (``hadronic noise") at $pp$ colliders. Furthermore, its large center-of-mass energy and luminosity can provide access to higher mass scales than most of the proposed $e^+e^-$ colliders. 
This capability of $e^- p$ colliders in the case of long-lived particles with soft decay products and/or very short lifetimes (up to $c \tau \sim \mu$m) was demonstrated in \cite{Curtin:2017bxr}. In the current study, we examine the potential of LHeC to observe a prompt signal with multiple soft jets, which looks like hadronic noise at $pp$ colliders thus making it very hard to observe. As mentioned in the previous section, such a signal can come from the prompt decays of light RPV gluinos.

\section{Probing Light RPV Gluinos at the LHeC}
\label{s.gluinosatep}

\subsection{Signal}

\begin{figure}
\centering
\includegraphics[scale=0.25]{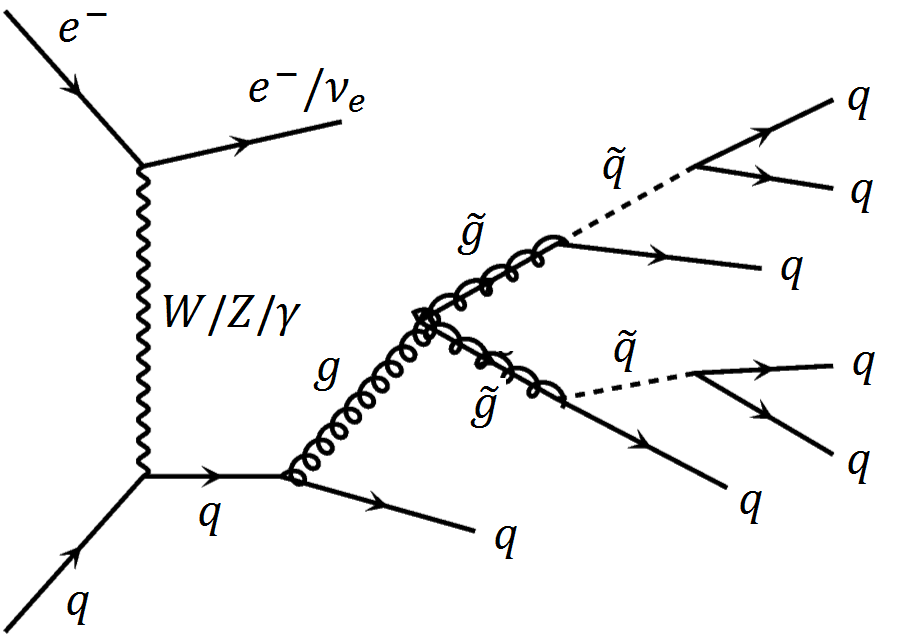}
\caption{Pair production of gluinos at the LHeC followed by RPV decay to $jjj$}
\label{fig:Feynman_diagram_RPV_gluino_signal}
\end{figure}

The production of RPV gluinos is simulated in MG5 aMC@NLO \cite{Alwall:2014hca} at parton-level using the RPVMSSM model \cite {Fuks:2012im}. We perform a simple parton-level study which is sufficient to robustly demonstrate sensitivity, leaving more sophisticated studies for future work.
In an $e^- p$ collision, gluinos are pair-produced mostly via the gluon radiated off from the initial or final state quark. Each gluino then undergoes RPV decay to $jjj$ via an intermediate heavy squark (see Fig.~\ref{fig:Feynman_diagram_RPV_gluino_signal}). 
The final state thus has 7 jets, including the jet from the incoming parton.

In this paper, we will consider two benchmark values of the gluino masses in the light gluino gap as discussed in Section~\ref{s.lightgluinos}: 50 and 70 GeV. Note that since the production cross section is fully determined by the quantum numbers of the gluino, there are no additional free parameters.\footnote{We will ignore the potential R-parity conserving decays of gluino into gravitino, but these will in any case be very small due to the existing bounds.}
The corresponding LHeC cross sections, for events with one (zero) electron in the final state, for these points are 38 (14) and 12 (4.7) fb at LO parton level, respectively. For the multi-jet final state from gluino decays, we use the following basic generator-level cuts for jet reconstruction: 
\begin{equation}
\label{eq:generator_level_cuts_on_jets}
p_T > 20 \ ~\rm{GeV}, \ \eta < 5, \ \Delta R > 0.4
\end{equation}
$\Delta R > 0.4$ is imposed on all possible pairs of jets in an event by manually merging the parton-level jets which are too close to each other. We remark that these are conservative choices which can be relaxed if needed;  for instance, one could use $p_T > 10$ GeV provided that $E > 20$ GeV, or cluster jets with smaller radius.

\begin{figure}[t]
\centering
\includegraphics[width=\linewidth]{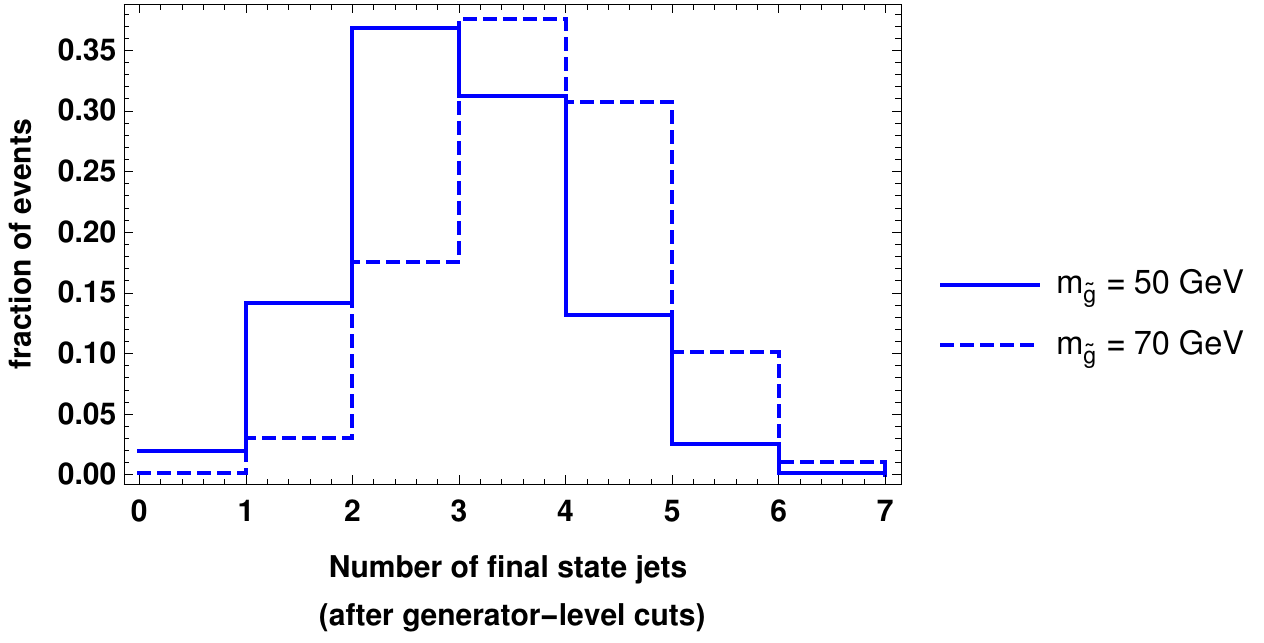}
\caption{Distributions for number of final state jets in signal events}
\label{fig:JetNumberDistribution}
\end{figure}

After applying these generator-level cuts, we are left with mostly 4-5 final state jets in an event, as can be seen from Fig. \ref{fig:JetNumberDistribution}. Hence, the 7-jet signal event from the gluino pair is effectively detected mostly as a 4- or 5-jet event. This is mostly a consequence of the simple assumptions on jet reconstruction we have made in our parton-level study -- a full detector-level analysis with more sophisticated jet substructure techniques would likely have access to much more information on the full jet multiplicity, making our reach estimates conservative. However, this reduced effective jet multiplicity also makes it possible to estimate the SM QCD background with standard Monte Carlo techniques, as we now discuss.


\subsection{Estimating SM background and signal significance}

\begin{figure*}
\centering
\includegraphics[width=\textwidth]{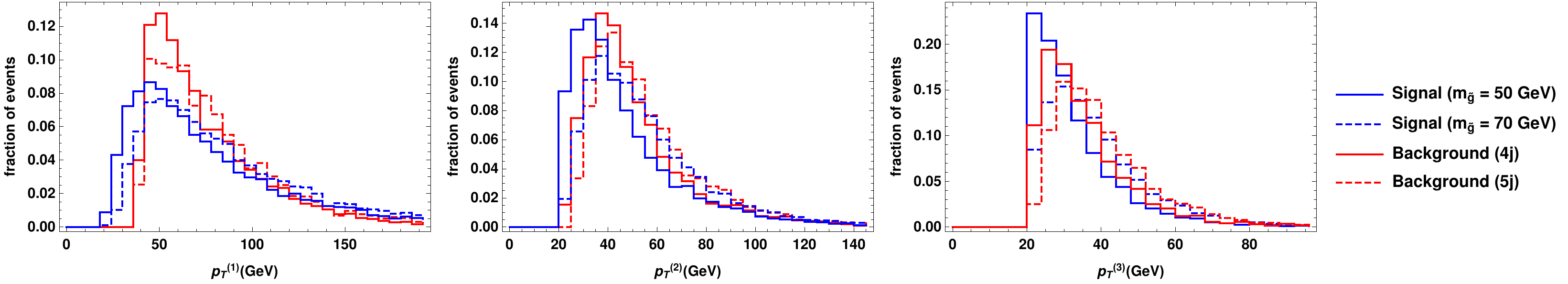}
\caption{$p_T$ distributions for the three hardest jets for the signal and background events (after applying the generator-level cuts in \eqref{eq:generator_level_cuts_on_jets}): The blue solid and dashed lines refer to the signal from $m_{\tilde{g}} = $ 50 and 70 GeV, respectively. The red solid and dashed lines refer to the QCD background with 4 and 5 jets, respectively.}
\label{fig:JetpTSignalBackground}
\end{figure*}

The dominant source of SM background for this multi-jet final state obviously comes from QCD processes. The tree-level estimates for cross section for these signal processes at the LHeC for one (zero) electron in the final state is 5.3 (0.27) pb and 0.58 (0.17) pb for 4-jet and 5-jet events, respectively. 
This is obtained at parton-level using the same generator-level cuts for jets as in \eqref{eq:generator_level_cuts_on_jets}. Looking at the $p_T$-distribution of the jets, both for the signal and QCD background events (see Fig.~\ref{fig:JetpTSignalBackground}), we consider using the signal selection cuts as defined in Table~\ref{table:signal_selection_cuts}
\footnote{The low $p_T$ threshold for jets at the LHeC of 10 GeV can be used to apply even more efficient signal selection cuts than here, based on the distributions in Fig.~\ref{fig:JetpTSignalBackground}.}
.

\begin{table}[h]
\centering
\begin{tabular}{|c|c|c|c|c|}
\hline
    	   & $p_T^{(1)}$(GeV) & $p_T^{(2)}$(GeV) & $p_T^{(3)}$(GeV) & $N_{jets}$ \\
\hline
Cut 1 &	 $\geq 50$ & $\geq 30$ & $\geq 25$ & $\geq 4$ \\
\hline
Cut 2 &	 $\geq 70$ & $\geq 40$ & $\geq 25$ & $\geq 4$ \\
\hline
Cut 3 &	 $\geq 70$ & $\geq 40$ & $\geq 25$ & $\geq 5$ \\
\hline
\end{tabular}
\caption{Signal selection cuts: $p_T^{(1)}, p_T^{(2)}, p_T^{(3)}$ are $p_T$ of jets ranked by their $p_T$. $N_{jets}$ is the number of jets in an event.}
\label{table:signal_selection_cuts}
\end{table}

\begin{figure*}
\centering
\includegraphics[width=\textwidth]{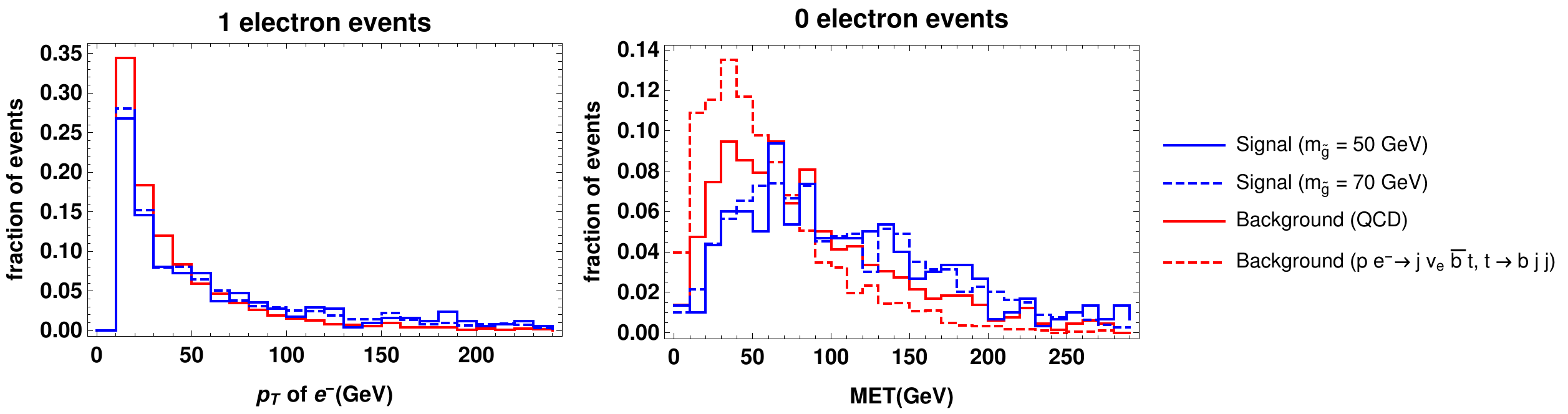}
\caption{Distributions for the $p_T$ of outgoing electron or MET for events with one and zero electron in the final state, respectively: These distributions are obtained after applying the Cut 3 (see Table~\ref{table:signal_selection_cuts}). The blue solid and dashed lines refer to the signal from $m_{\tilde{g}} = $ 50 and 70 GeV, respectively. The red solid line shows the QCD background. The red dashed line is the background contribution to zero electron events from 
$p + e^- \rightarrow j + \nu_e + \bar{b} + t , t \rightarrow b + jj$, which has cross section comparable to that of the QCD processes after applying Cut 3 (see Table~\ref{table:XSall}).}
\label{fig:pT_electron_MET_Signal_Background}
\end{figure*}

\begin{figure*}
\centering
\includegraphics[width=\textwidth]{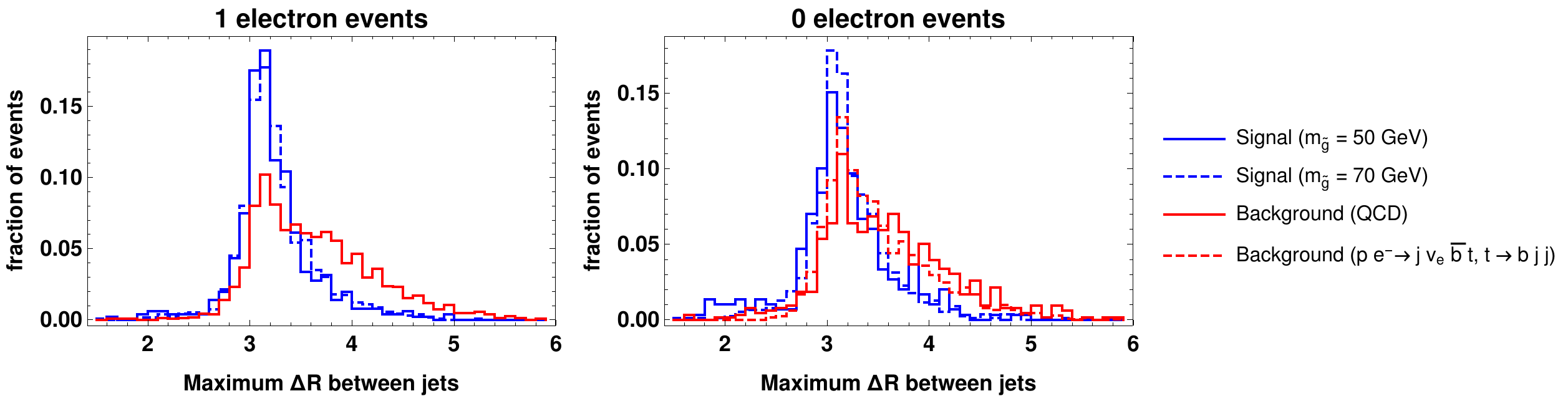}
\caption{Distributions for the maximum $\Delta R$ among all pairs of jets in an event: These distributions are obtained after applying the Cut 3 (see Table~\ref{table:signal_selection_cuts}). The blue solid and dashed lines refer to the signal from $m_{\tilde{g}} = $ 50 and 70 GeV, respectively. The red solid line shows the QCD background. The red dashed line is the background contribution to zero electron events from 
$p + e^- \rightarrow j + \nu_e + \bar{b} + t , t \rightarrow b + jj$, which has cross section comparable to that of the QCD processes after applying Cut 3 (see Table~\ref{table:XSall}).}
\label{fig:DeltaR_jet_max_Signal_Background}
\end{figure*}

\begin{table*}
\centering
\begin{tabular}{|c|c|c|c|c|c|}
\hline
  & Generator-level & Cut 1 & Cut 2 & Cut 3 & \makecell{Cut 3 + \\ $\ p_T(e^-)/{\rm MET}$ + $\Delta R^{jet}_{max}$} \\
\hline
Signal for $m_{\tilde{g}} = 50$ GeV & 38 (14) & 11.13 (5.47) & 6.52 (3.90) & 2.61 (1.60) & 0.29 (0.36)\\
Signal for $m_{\tilde{g}} = 70$ GeV & 12 (4.7) & 6.78 (3.33) & 4.23 (2.48) & 2.33 (1.33) & 0.20 (0.27) \\
\hline
\multirow{2}{*}{QCD background} & 4j: 5270 (270) & \multirow{2}{*}{3140 (350)} & \multirow{2}{*}{1610 (230)} & \multirow{2}{*}{238 (96)} & \multirow{2}{*}{6.29 (8.63)} \\
& 5j: 579 (169) & & & & \\
\hline
$p+e^- \rightarrow j+e^-/\nu_e+j+V, V \rightarrow jj$ & 230 (157) & 131 (101) & 88 (74) & 0 (0) & 0 (0) \\
$p + e^- \rightarrow j + e^-/\nu_e + b\bar{t}/\bar{b}t , t/\bar{t} \rightarrow b/\bar{b} + jj$ & 0.1 (171) & 0.09 (131) & 0.08 (92) & 0.04 (42) & $<0.01$ (1.24) \\
$p + e^- \rightarrow j + e^-/\nu_e + j + h, h \rightarrow b \bar{b}$ & 1.5 (7.5) & 1.2 (6.0) & 0.98 (4.8) & 0 (0) & 0 (0)\\
$p + e^- \rightarrow j + e^-/\nu_e + VV, V \rightarrow jj$ & 
5.2 (3.6) & 4.4 (3.2) & 3.6 (2.8) & 1.9 (1.5) & $<0.1$ ($<0.1$)\\
$p + e^- \rightarrow j + e^-/\nu_e + t\bar{t}, t\bar{t} \rightarrow$ hadronic & 1.96 (0.01) & 1.9 (0.01) & 1.82 (0.01) & 1.77 (0.01) & $<0.01$ ($<0.01$)\\
$p + e^- \rightarrow j + e^-/\nu_e + t\bar{t}, t\bar{t} \rightarrow$ semi-leptonic & 1.96 (0.01) & 1.83 (0.01) & 1.6 (0.01) & 0.99 (0.01) & $<0.01$ ($<0.01$)\\
\hline
\multirow{2}{*}{Total background} & 4j: 5500 (435) & \multirow{2}{*}{3280 (590)} & \multirow{2}{*}{1710 (402)} & \multirow{2}{*}{242 (139)} & \multirow{2}{*}{6.3 (9.9)} \\
& 5j: 708 (224) & & & & \\
\hline
\end{tabular}
\caption{Cross sections (in fb) for the signal and various SM background processes for one (zero) electron in the final state: $V$ refers to all the electroweak gauge bosons. The Cuts 1-3 are as defined in Table~\ref{table:signal_selection_cuts}. In the last column, the additional cuts are as defined in \eqref{eq:additional_cuts_for_SysError} with $p_T^{min} = 140$ GeV and $\Delta R^* = 3.2$.}
\label{table:XSall}
\end{table*}

\begin{table*}
\centering
\begin{tabular}{|c|c|c|c|c|c|}
	\hline
	  & $m_{\tilde{g}}$ (GeV) & Cut 1 & Cut 2 & Cut 3 & \makecell{Cut 3 + \\ $\ p_T(e^-)/{\rm MET}$ + $\Delta R^{jet}_{max}$} \\
	\hline
	\multirow{2}{*}{$S/B \times 100$} & 50 & (0.34, 0.92) & (0.38, 0.97) & (1.1, 1.1) & (4.6, 3.6) \\
	\cline{2-6}
	&  70 & (0.21, 0.56) & (0.25, 0.62) & (0.96, 0.95) & (3.2, 2.7) \\
	\hline
	\multirow{4}{*}{Statistical significance, $S/\sqrt{B}$} & 50 & \makecell{9.4 \\ (6.15, 7.11)} & \makecell{7.91 \\ (4.98, 6.15)} & \makecell{6.92 \\ (5.30, 4.29)} & \makecell{5.12 \\ (3.63, 3.61)} \\
	\cline{2-6}
	&  70 & \makecell{5.72 \\ (3.74, 4.33)} & \makecell{5.07 \\ (3.23, 3.91)} & \makecell{5.92 \\ (4.73, 3.56)} & \makecell{3.74 \\ (2.54, 2.74)} \\
	\hline
	\multirow{4}{*}{Significance with $1\%$ systematic error} & 50 & \makecell{0.98 \\ (0.34, 0.92)} & \makecell{1.03 \\ (0.38, 0.96)} & \makecell{1.53 \\ (1.05, 1.11)} & \makecell{3.83 \\ (2.84, 2.56)} \\
	\cline{2-6}
	 & 70 & \makecell{0.60 \\ (0.21, 0.56)} & \makecell{0.67 \\ (0.25, 0.61)} & \makecell{1.31 \\ (0.94, 0.92)} & \makecell{2.78 \\ (1.99, 1.94)} \\
	\hline
\end{tabular}
\caption{$S/B$ and signal significance for RPV gluino signal events at the LHeC for $m_{\tilde{g}} = 50, 70$ GeV: Signal significance with systematic error is computed using \eqref{eq:signal_significance_with_SysError}. The Cuts 1-3 are as defined in Table~\ref{table:signal_selection_cuts}. In the last column, the additional cuts are as defined in \eqref{eq:additional_cuts_for_SysError} with $p_T^{min} = 140$ GeV and $\Delta R^* = 3.2$. Signal significance for one and zero electron events, written inside the brackets in the same order, are added in quadrature to get the combined signal significance.}
\label{table:Significance}
\end{table*}

The signal and SM background cross sections after applying the cuts from Table~\ref{table:signal_selection_cuts} are listed in Table~\ref{table:XSall}, separately for events with one and zero electron in the final state. As expected, the non-QCD SM background processes are quite sub-dominant compared to the QCD processes.
We compute signal significance for the two types of events, with one and zero electron in the final state, separately as $\sigma_{1 e}$ and $\sigma_{0 e}$. The combined signal significance is obtained by adding these two in quadrature: $\sigma_{net} = \sqrt{\sigma_{1 e}^2 + \sigma_{0 e}^2}$. 
The combined signal significance neglecting systematic effects ($S/\sqrt{B}$) of $\geq 5 \sigma$ can be achieved both for 50 and 70 GeV gluinos using the above-mentioned cuts, as can be seen from Table~\ref{table:Significance}.  
Hence, in the absence of any systematic error, these light RPV gluinos can manifest themselves via excesses in the multi-jet events with $\geq 5 \sigma$ significance at the LHeC.

We now analyze the robustness of this search once systematic errors are taken into account. A 1\% systematic error is an achievable and reasonable goal, given the proposed design of the LHeC detector \cite{AbelleiraFernandez:2012cc}, the possibility of data-driven background estimates in a full analysis, and progress on theoretical predictions for backgrounds. 
We consider the following simplified expression for signal significance in the presence of a systematic error $(\lambda)$,
\begin{equation}
\label{eq:signal_significance_with_SysError}
\textrm{Significance} = \frac{S}{\sqrt{B + (\lambda B)^2}}
\end{equation}
with $\lambda = 1\%$. As can be seen from Table~\ref{table:Significance}, the signal significance is greatly reduced after taking into account the systematic error. For instance for the Cut 3 which is the strongest cut, the significance goes down from $\gtrsim 6\sigma$ to $\lesssim 1.5\sigma$, both for 50 and 70 GeV gluino mass. Hence, this demands for additional signal selection strategy to incorporate for the systematic error.

The signal processes involve harder momentum transfer than the corresponding QCD processes which form the dominant SM background. Hence the $p_T$ imparted to the outgoing $e^- / \nu_e$ in the signal processes is slightly greater than that of the background. (See Fig~\ref{fig:pT_electron_MET_Signal_Background}.) 
Also, the signal jets are expected to be more isotropic than the QCD background jets, since they are generated from gluino decays. This implies that the maximum angular separation between a pair of jets in a signal event will be slightly less than that in a background event. (See Fig~\ref{fig:DeltaR_jet_max_Signal_Background}.) 
Following these simple observations, we can employ the following additional cuts: 
\begin{equation}
\label{eq:additional_cuts_for_SysError}
p_T(e^-) \textrm{ or }  {\rm MET} > p_T^{min} \ ; \ \Delta R^{jet}_{max} < \Delta R^*
\end{equation}
The first cut corresponds to $p_T(e^-) > p_T^{min}$ for events with one $e^-$ in the final state and ${\rm MET} > p_T^{min}$ for events with no $e^-$ in the final state. The second cut refers to the maximum $\Delta R$ among all pairs of jets in an event being less than $\Delta R^*$.
 
The optimal use of these cuts can be made with $p_T^{min} \approx 140$ GeV and $\Delta R^* \approx 3.2$. The cross sections for one (zero) electron events after applying these cuts are as follows: 6.3 (9.9) fb for the background, 0.29 (0.36) fb for the signal from $m_{\tilde{g}} = 50$ GeV, 0.20 (0.27) fb for the signal from $m_{\tilde{g}} = 70$ GeV. As listed in Table~\ref{table:Significance}, this gives the combined signal significance of $\approx 3.8\sigma$ and $\approx 2.8\sigma$ for $m_{\tilde{g}} =$ 50 and 70 GeV, respectively. 
The signal significance for $m_{\tilde{g}} =$ 70 GeV is lower than that of $m_{\tilde{g}} =$ 50 GeV by a factor of $\sim$0.7. This is a combination of the production cross section for the former being less by a factor of $\sim$0.3 (see Table~\ref{table:XSall}) and the $p_T$-based cuts being comparatively harder on the latter (see Fig.~\ref{fig:JetpTSignalBackground} and Table~\ref{table:XSall}).
Thus, this shows that even in the presence of $1 \%$ systematic error, we can use simple cuts like \eqref{eq:additional_cuts_for_SysError} and reach 2$\sigma$ exclusion limit for this signal at the LHeC. More sophisticated signal selection strategies based on jet sub-structure, machine-learning techniques and with more knowledge of the detector are expected to significantly enhance the prospects of signal detection.

\section{Conclusions}
\label{s.conclusions}

RPV or Stealth SUSY can give rise to gluino with all-hadronic and multi-jet decay with no missing energy signal. This is not subject to constraints from the standard SUSY searches at the LHC. As pointed out in \cite{Evans:2018scg}, there is an existing gap in the search for such light gluinos with $m_{\tilde{g}} \approx 50-70$ GeV. Such a BSM scenario giving a soft multi-jet signal is very challenging to probe at the LHC (or future hadron colliders) due to the large pile-up and background. 
In this work, for the benchmark case of RPV SUSY, we have shown that this gap can be covered at the LHeC. Even in the presence of $1 \%$ systematic error, using simple signal selection cuts based on the $p_T$ of jets, $e^-$ and the isotropy of signal jets, the LHeC has the potential to observe soft multi-jet signal from these light RPV gluinos with $\geq$ 2$\sigma$ significance i.e. at 95 $\%$ confidence level.  

We demonstrated that the LHeC has a unique sensitivity to BSM scenarios giving soft multi-jet signals which are very hard to probe at the $pp$ colliders, such as the here considered RPV gluino. It is likely that similar opportunities exist for hidden valley models \cite{Strassler:2006im,Strassler:2006ri,Strassler:2006qa,Han:2007ae,Strassler:2008bv,Strassler:2008fv} which can yield high-multiplcity soft hadronic signals at detectable rates. Thus, as first illustrated in \cite{Curtin:2017bxr}, $e^-p$ colliders have a unique potential to detect BSM signals that look like hadronic noise at the $pp$ colliders thus providing a complementary discovery potential to the future $pp$ colliders. Further study is required to identify more such generic BSM scenarios which will add towards the motivation for the construction of the future proposed $e^-p$ colliders.


\acknowledgments
We thank the LHeC working group for their feedback and valuable discussions. 
We thank L. Delle Rose for collaboration in the early stages of this work.
K.D. is supported by National Science Foundation grant No. PHY-1620074 and the Maryland Center for Fundamental Physics. 
O.F.\ has received funding from the European Unions Horizon 2020 research and innovation program under the Marie Sklodowska-Curie grant agreement No 674896 (Elusives).

\bibliography{ep_collider}

\end{document}